\documentclass[letterpaper,12pt]{article}
\usepackage{array}
\usepackage{dcolumn}
\usepackage{float}
\usepackage{natbib}
\usepackage{graphicx}
\usepackage{color}
\usepackage{amsmath}
\usepackage{amsfonts}
\usepackage{amsopn}
\usepackage{amsthm}
\usepackage{lscape}
\usepackage{multirow}
\bibpunct{}{}{,}{s}{,}{,}
\newtheorem{thm}{Theorem}

\newtheorem{prop}[thm]{Proposition}
\theoremstyle{definition}

\theoremstyle{remark}

\DeclareMathOperator{\pr}{P}
\DeclareMathOperator{\epn}{E}
\DeclareMathOperator{\var}{Var}
\DeclareMathOperator{\cov}{Cov}

\DeclareMathOperator{\argmin}{arg\;min}
\newcommand{\bmX}{\boldsymbol X}
\newcommand{\bmZ}{\boldsymbol Z}
\newcommand{\bmW}{\boldsymbol W}
\newcommand{\bmO}{\boldsymbol O}
\newcommand{\bmV}{\boldsymbol V}
\newcommand{\bmzero}{\boldsymbol 0}
\newcommand{\bmbeta}{\mbox{\boldmath${\beta}$}}

\renewcommand{\baselinestretch}{1.8}

\begin{document}
\begin{center}
{\LARGE A Connection Between Covariate Adjustment and Stratified Randomization in Randomized Clinical Trials}
\end{center}

\vspace{0.3cm}
\begin{center}
Zhiwei Zhang\\
Biostatistics Innovation Group, Gilead Sciences, Foster City, California, USA\\
zhiwei.zhang6@gilead.com
\end{center}

\vspace{0.3cm}
\centerline{\bf Summary}

The statistical efficiency of randomized clinical trials can be improved by incorporating information from baseline covariates (i.e., pre-treatment patient characteristics). This can be done in the design stage using stratified (permutated block) randomization or in the analysis stage through covariate adjustment. This article makes a connection between covariate adjustment and stratified randomization in a general framework where all regular, asymptotically linear estimators are identified as augmented estimators. From a geometric perspective, covariate adjustment can be viewed as an attempt to approximate the optimal augmentation function, and stratified randomization improves a given approximation by moving it closer to the optimal augmentation function. The efficiency benefit of stratified randomization is asymptotically equivalent to attaching an optimal augmentation term based on the stratification factor. In designing a trial with stratified randomization, it is not essential to include all important covariates in the stratification, because their prognostic information can be incorporated through covariate adjustment. Under stratified randomization, adjusting for the stratification factor only in data analysis is not expected to improve efficiency, and the key to efficient estimation is incorporating prognostic information from all important covariates. These observations are confirmed in a simulation study and illustrated using real clinical trial data.

\noindent{Key words:}
analysis of covariance; augmented estimator; covariate-adaptive randomization; simple randomization; standardized logistic regression; super learner

\section{Introduction}\label{intro}

Randomized clinical trials are widely considered the gold standard for evaluating the safety and effectiveness of medical treatments. It is well recognized that baseline covariates (i.e., pre-treatment patient characteristics) can be used to improve the statistical efficiency of treatment effect estimation in the analysis of trial data. This usage of baseline covariates, commonly known as covariate adjustment, has received a great deal of attention from practitioners and regulators \citep{fda23}. Another common usage of baseline covariates in clinical trials is stratified (permuted block) randomization \citep{z74}, which aims to improve the balance of selected baseline covariates between treatment groups. In this work, we study the relationship between covariate adjustment and stratified randomization with a focus on statistical efficiency.

There is a large and growing literature on statistical methods for covariate adjustment in clinical trials \citep{t08,z08,r08,m09,r10,t12,w19,zm19,z20,b21}. Under simple randomization, the semiparametric theory \citep{t06} can be used to characterize all regular, asymptotically linear estimators of common treatment effect measures such as the mean (risk) difference, the mean (risk) ratio, and the odds ratio. This characterization has motivated an augmentation approach in which an unadjusted estimator or estimating function, which does not involve covariate data, is augmented with a term based on treatment and an arbitrary function of covariates \citep{t08,z08}. The optimal augmentation, which minimizes the asymptotic variance of an augmented estimator, has also been characterized. To take advantage of this optimality result, one may use a working regression model or a machine learning algorithm to estimate the optimal augmentation and use the estimated augmentation to construct an augmented estimator. 
The methods and results referenced in this paragraph generally assume simple randomization; they are not immediately applicable to stratified randomization.

Stratified randomization is a special case of covariate-adaptive randomization (CAR) \citep{rs08,rl15}. Under CAR, treatment assignments for different subjects may depend on each other, and this dependence needs to be taken into account when analyzing treatment effect estimators. In recent years, considerable progress has been made in understanding the asymptotic properties of treatment effect estimators and test statistics under CAR \citep{s10,b18,ys20,yys22,y22,w23,b23}. While some of these references pertain to specific estimators, the work of \citet{w23} covers a large class of M-estimators (i.e., estimators that can be regarded as solutions to estimating equations), and the work of \citet{b23} covers the entire class of augmented estimators for common treatment effect measures. Loosely speaking, their results indicate that a treatment effect estimator that is asymptotically linear under simple randomization typically remains so under CAR, and its asymptotic variance may become smaller. Among the various forms of CAR studied, stratified randomization and the biased coin design \citep{e71} achieve the largest variance reduction because of their \lq\lq strong balance" property \citep{b18,yys22,b23}. In light of its statistical efficiency, simplicity and widespread use in practice, we will focus on stratified randomization as a representative of CAR.

In this article, we make a general connection between covariate adjustment and stratified randomization and discuss its statistical and practical implications. Under the aforementioned augmentation approach, which covers virtually all estimators of practical interest, we note that covariate adjustment can be viewed geometrically as an attempt to approximate the optimal augmentation function in a suitable function space. From this geometric perspective, stratified randomization can be interpreted as a way to improve a given approximation by projecting it into a certain subspace containing the optimal augmentation function. The efficiency benefit of stratified randomization is asymptotically equivalent to adding an optimal augmentation term based on the stratification factor. This connection helps characterize efficient estimators under stratified randomization and has important practical implications:
\begin{itemize}
\item In designing a trial with stratified randomization, there is often a conflict between the desire to include all important covariates in the stratification and the need to avoid having too many strata, some of which may be quite small. In light of the connection between stratified randomization and covariate adjustment, it is not essential to include all important covariates in the stratification, as covariate adjustment provides another opportunity to make use of their prognostic information. This flexibility can be helpful in practice, especially for covariates that are time-consuming to measure.
\item Under stratified randomization, adjusting for the stratification factor only in the analysis of trial data is not expected to improve statistical efficiency. Adjusting for the stratification factor is common practice in clinical trials and consistent with regulatory guidelines \citep{ich98,ema03,ema15,fda23}. However, the key to efficient estimation is incorporating prognostic information from all important baseline covariates. Such covariates may include continuous covariates underlying the stratification factor as well as additional covariates unrelated to the stratification factor.
\end{itemize}

To complement the above considerations based on asymptotic theory, we report a simulation study comparing different estimation methods under simple versus stratified randomization. Our simulation results confirm that adjusting for the stratification factor alone does not improve efficiency under stratified randomization, that large efficiency gains can be obtained by incorporating other covariates than the stratification factor, and that stratified randomization makes little difference to estimators that adjust for all covariates. Among the different methods for covariate adjustment, a nonparametric method based on the super learner \citep{zm19} performs similarly to or better than standard methods based on parametric models, depending on how well the parametric models approximate the truth. 

The rest of the article is organized as follows. In Section \ref{sec:prelim}, we formulate the estimation problem, characterize covariate-adjusted treatment effect estimators, and summarize asymptotic results for stratified randomization. In Section \ref{sec:rst.impl}, we make a connection between covariate adjustment and stratified randomization, and discuss its statistical and practical implications. A simulation study is reported in Section \ref{sim}, followed by a real example in Section \ref{example}. The article concludes with a discussion in Section \ref{disc}. Appendix A describes a specific type of covariate-adjusted estimators and Appendix B provides proofs of theoretical results.

\section{Preliminaries}\label{sec:prelim}

\subsection{Estimation Problem}\label{sec:estimand}

We start by formulating the estimation problem using potential outcomes \citep{r74}. For a generic patient in the target population, let $Y(a)$ denote the potential outcome for treatment $a$, where $a=1$ for an experimental treatment and $a=0$ for a control treatment, which may be placebo or standard of care. For each $a\in\{0,1\}$, we assume $\epn\{Y(a)^2\}<\infty$ and denote $\mu_a=\epn\{Y(a)\}$. Suppose the treatment effect of interest is $\delta=g(\mu_1)-g(\mu_0)$, where $g$ is a smooth and strictly increasing function specified by the investigator. Popular choices for $g$ include identity (for continuous and other types of outcomes), logarithm (for outcomes with positive means) and logit (for binary outcomes). 

Let $\bmW$ be a vector of available baseline covariates (i.e., pre-treatment patient characteristics) that may be associated with either or both of $Y(1)$ and $Y(0)$. Let $\bmZ_i=(Y_i(1),Y_i(0),\bmW_i)$, $i=1,\dots,n$, be independent copies of $\bmZ=(Y(1),Y(0),\bmW)$ arising from a random sample of $n$ subjects enrolled in a clinical trial. For the $i$th subject in the trial, let $A_i$ denote the randomized treatment assignment (to be discussed later) and $Y_i$ the actual observed outcome. Under the assumption of consistency or stable unit treatment value, we have $Y_i=Y_i(A_i)=A_iY_i(1)+(1-A_i)Y_i(0)$, $i=1,\dots,n$. The observed data will be denoted by $\bmO_i=(\bmW_i,A_i,Y_i)$, $i=1,\dots,n$. 

\subsection{Covariate Adjustment}\label{sec:cov.adj}

While stratified randomization is of particular interest in this article, simple randomization provides a natural framework for introducing estimators. Therefore, in the rest of this subsection, we assume that simple randomization is used to make treatment assignments. This means that the $A_i$'s are independent of each other and of the $\bmZ_i$'s, with $\pr(A_i=1)\equiv\pi\in(0,1)$. It follows that the $\bmO_i$'s are independent copies of $\bmO=(\bmW,A,Y)$, where $A$ is Bernoulli with $\pr(A=1|\bmZ)=\pr(A=1)=\pi$ and $Y=Y(A)=AY(1)+(1-A)Y(0)$.

It is straightforward to estimate $\delta$ with the empirical estimator $\widehat\delta_{\text{emp}}=g\left(\overline Y_1\right)-g\left(\overline Y_0\right)$, where $\overline Y_a=\{\sum_{i=1}^nI(A_i=a)\}^{-1}\sum_{i=1}^nI(A_i=a)Y_i$, $a=0,1$, and $I(\cdot)$ is the indicator function. It is well known that $\widehat\delta_{\text{emp}}$ is consistent and asymptotically normal in the sense that $\sqrt n(\widehat\delta_{\text{emp}}-\delta)$ converges to a zero-mean normal distribution. Its (scaled) asymptotic variance is given by $\var\{\psi(\bmO)\}$, where
\begin{equation*}\label{psi.marg.effect}
\psi(\bmO)=g'(\mu_1)\frac{A(Y-\mu_1)}{\pi}-
g'(\mu_0)\frac{(1-A)(Y-\mu_0)}{1-\pi}
\end{equation*}
and $g'$ is the derivative function of $g$. The empirical estimator may be inefficient as it does not make use of the available information in $\bmW$.

The information in $\bmW$ can be incorporated using an augmentation approach \citep{t08,z08}. In the present setting, an augmented estimator of $\delta$ may be obtained as $\widehat\delta_{\text{aug}}(b)=\widehat\delta_{\text{emp}}-n^{-1}\sum_{i=1}^n(A_i-\pi)b(\bmW_i)$, where $b(\bmW)$ is a real-valued function of $\bmW$ such that $\|b\|_2^2:=\epn\{b(\bmW)^2\}<\infty$. The set of all such functions is denoted by $\mathcal B$ and regarded as an $L_2$ space. For any fixed $b\in\mathcal B$, $\widehat\delta_{\text{aug}}(b)$ is consistent and asymptotically normal with asymptotic variance $\sigma^2(b)=\var\{\psi(\bmO)-(A-\pi)b(\bmW)\}$. Furthermore, any estimator of $\delta$ that is regular and asymptotically linear must be asymptotically equivalent to $\widehat\delta_{\text{aug}}(b)$ for some $b$ (and thus have the same asymptotic variance) \citep{t06}. The optimal choice of $b$, which minimizes $\sigma^2(b)$, is given by
$$
b_{\text{opt}}(\bmW)=\frac{g'(\mu_1)\{m_1(\bmW)-\mu_1\}}{\pi}+\frac{g'(\mu_0)\{m_0(\bmW)-\mu_0\}}{1-\pi},
$$
where $m_a(\bmW)=\epn(Y|\bmW,A=a)=\epn\{Y(a)|\bmW\}$, $a=0,1$.

Although $b_{\text{opt}}$ is unknown, it can be estimated from trial data using a regression model for $\epn(Y|A,\bmW)$ \citep{t08} or a machine learning algorithm \citep{zm19}. Let $\widehat b$ be a generic estimator of $b_{\text{opt}}$ and assume that $\widehat b$ converges to some limit function $b\in\mathcal B$, which may be different from $b_{\text{opt}}$, in the sense that $\|\widehat b-b\|_2=o_p(1)$. Under quite general conditions, $\widehat\delta_{\text{aug}}(\widehat b)$ is asymptotically equivalent to $\widehat\delta_{\text{aug}}(b)$ with the same asymptotic variance $\sigma^2(b)$. This is the case if $\widehat b$ belongs to a Donsker class \citep{vw96} or if $\widehat\delta_{\text{aug}}(\widehat b)$ employs sample splitting \citep{z11,c18,k20}. We will not be concerned about these technical issues in this article and will simply assume that $\widehat b$ and $\widehat\delta_{\text{aug}}(\widehat b)$ possess the properties stated in this paragraph.

In Appendix A, we describe a class of regression estimators of $\delta$ that are commonly used in practice. These regression estimators are not constructed as, but are asymptotically equivalent to, augmented estimators.

\subsection{Stratified Randomization}\label{sec:strat}

Stratified randomization involves stratifying the patient population on the basis of one or more baseline covariates, which we assume are included in $\bmW$. Let the support of $\bmW$ be partitioned as $\mathcal W=\cup_{k=1}^K\mathcal W_k$ with $\mathcal W_{k_1}\cap\mathcal W_{k_2}=\emptyset$ for $k_1\not=k_2$. Let $S=s(\bmW)=\sum_{k=1}^KkI(\bmW\in\mathcal W_k)$, and let $S_i=s(\bmX_i)$, $i=1,\dots,n$. While simple randomization would be sufficient to ensure that $\pr(A=1|S)=\pr(A=1)=\pi$, stratified randomization aims to achieve better balance within each stratum defined by $S$. It does so by randomizing subjects in consecutive blocks within each stratum, with the constraint that the proportion of subjects with $A_i=1$ must be equal to $\pi$ within each completed block. Under stratified randomization, it is evident that $(A_1,\dots,A_n)$ are conditionally independent of $(\bmZ_1,\dots,\bmZ_n)$ given $(S_1,\dots,S_n)$. The observed data, $\{\bmO_i=(\bmW_i,A_i,Y_i),i=1,\dots,n\}$, are no longer independent across subjects because the $A_i$'s may depend on each other. On the other hand, the $\bmO_i$'s are identically distributed and their (common) marginal distribution is the same as the distribution of $\bmO=(\bmW,A,Y)$ (described at the beginning of Section \ref{sec:cov.adj}).

It is straightforward to deduce the asymptotic properties of augmented estimators from the recent theoretical work on CAR \citep{b18,w23,b23}. As before, suppose $\widehat b$ is an estimator of $b_{\text{opt}}$ that converges to some fixed function $b$. Under stratified randomization, the augmented estimators $\widehat\delta_{\text{aug}}(b)$ and $\widehat\delta_{\text{aug}}(\widehat b)$ are consistent and asymptotically normal with the same asymptotic variance, which we denote by $\sigma_{\text{st}}^2(b)$; that is, both $\sqrt n\{\widehat\delta_{\text{aug}}(b)-\delta\}$ and $\sqrt n\{\widehat\delta_{\text{aug}}(\widehat b)-\delta\}$ converge to the same normal distribution with mean 0 and variance $\sigma_{\text{st}}^2(b)$. \citet{w23} give the following characterization of $\sigma_{\text{st}}^2(b)$:
\begin{equation}\label{strat.var.w23}
\sigma_{\text{st}}^2(b)=\sigma^2(b)
-\frac{1}{\pi(1-\pi)}\epn\left\{\left(\epn[(A-\pi)\{\psi(\bmO)-(A-\pi)b(\bmW)\}|S]\right)^2\right\}.
\end{equation}
Recall that $\sigma^2(b)$ is the common asymptotic variance of $\widehat\delta(b)$ and $\widehat\delta(\widehat b)$ under simple randomization. Clearly, $\sigma_{\text{st}}^2(b)\le\sigma^2(b)$, where equality holds if and only if $\epn[(A-\pi)\{\psi(\bmO)-(A-\pi)b(\bmW)\}|S]\equiv0$. Thus, compared to simple randomization, stratified randomization does not increase, and typically decreases, the asymptotic variance of an augmented estimator.

\section{Results and Implications}\label{sec:rst.impl}

\subsection{Geometric Interpretations}\label{sec:geo}

In this subsection, we make a connection between covariate adjustment and stratified randomization based on their geometric interpretations. The geometric arguments we use to make the connection are helpful but not essential for understanding the statistical and practical implications of the connection (to be discussed in Sections \ref{sec:stat.impl} and \ref{sec:prac.impl}).

We note first that covariate adjustment can be viewed geometrically as an attempt to approximate the optimal augmentation function $b_{\text{opt}}$. Not only do we know that $b_{\text{opt}}$ minimizes the asymptotic variance $\sigma^2(b)$ of an augmented estimator, we can show that the value of $\sigma^2(b)$ is directly related to the mean square (i.e., $L_2$) distance between $b$ and $b_{\text{opt}}$.

\begin{prop}\label{rst:geo.cv}
For any $b\in\mathcal B$, we have
\begin{equation}\label{geo.cv}
\sigma^2(b)=\sigma^2(b_{\text{opt}})+\pi(1-\pi)\|b-b_{\text{opt}}\|_2^2.
\end{equation}
\end{prop}

\noindent In words, the extra asymptotic variance of a sub-optimal estimator, $\sigma^2(b)-\sigma^2(b_{\text{opt}})$, is proportional to the approximation error $\|b-b_{\text{opt}}\|_2^2$. Proposition \ref{rst:geo.cv} indicates that, while it may be unrealistic to hope that $b=b_{\text{opt}}$, it is generally helpful to reduce the approximation error $\|b-b_{\text{opt}}\|_2^2$.

\begin{figure}[H]
	\linespread{1.2}
    \centering
    \includegraphics[width=0.9\linewidth]{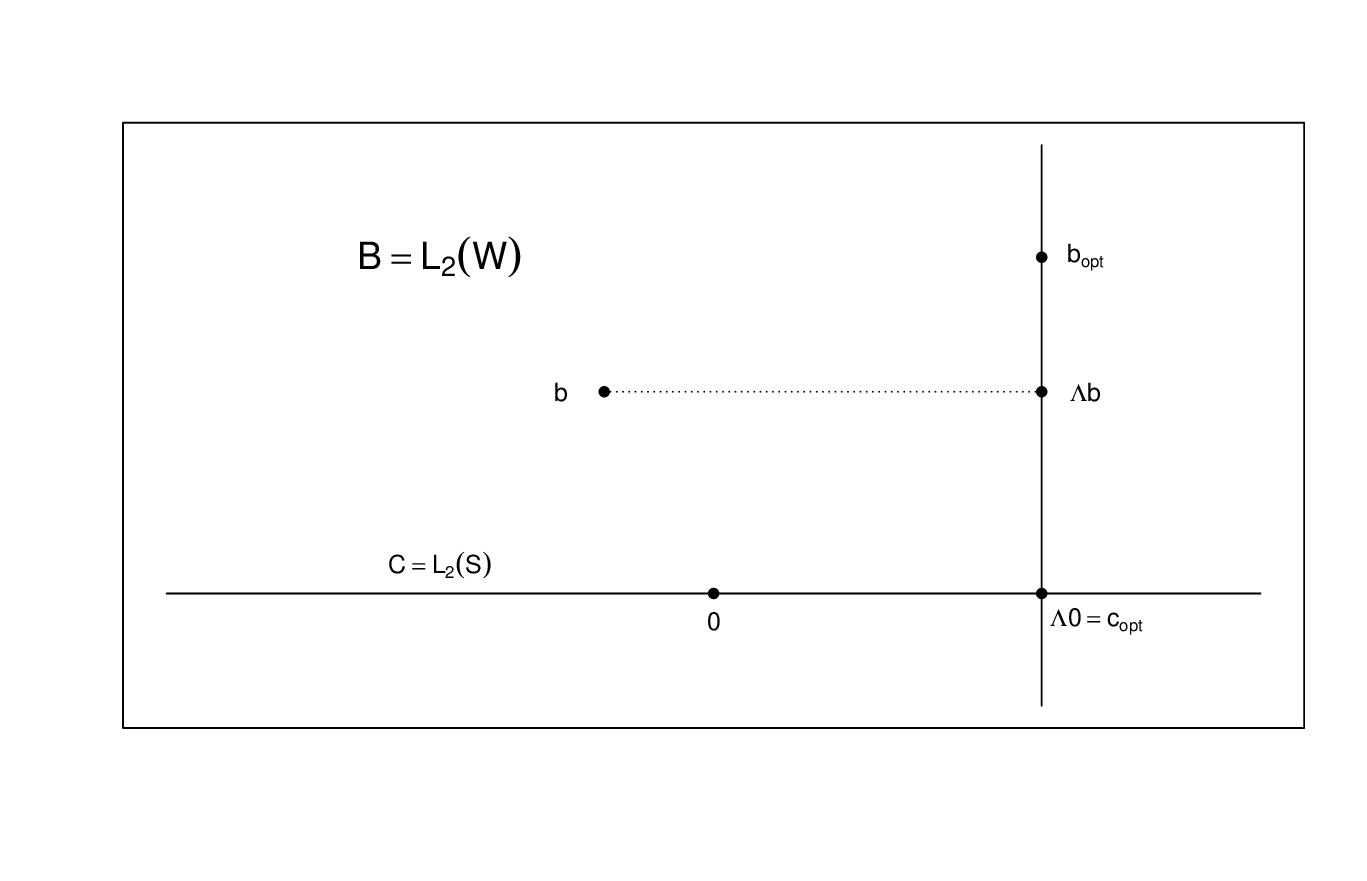}
    \caption{A geometric interpretation of stratified randomization. The plane represents $\mathcal B$, the space of augmentation functions. The horizontal line passing through 0 represents the subspace $\mathcal C$ consisting of functions of $S$. The vertical line passing through $b_{\text{opt}}$ represents the affine subspace $b_{\text{opt}}+\mathcal C^{\perp}$. The dotted horizontal line segment represents the operator $\Lambda$, which projects a given $b\in\mathcal B$ into the affine space $b_{\text{opt}}+\mathcal C^{\perp}$. The fact that $\Lambda b$ is closer to $b_{\text{opt}}$ than $b$ is explains the variance reduction effect of stratified randomization. The point $\Lambda 0=c_{\text{opt}}$ is the optimal augmentation function based on $S$. See Section \ref{sec:geo} for more details.}
    \label{fig:geo.strat}
\end{figure}

Next, we give a geometric interpretation to equation \eqref{strat.var.w23} that helps make a connection with covariate adjustment. We will show that the variance reduction effect of stratified randomization is equivalent to moving a given augmentation function $b$ closer to $b_{\text{opt}}$ under simple randomization. Some notations are necessary for describing this result. Recall that $\mathcal B$, the space of augmentation functions, is a Hilbert space with inner product $\langle b_1,b_2\rangle=\epn\{b_1(\bmW)b_2(\bmW)\}$. Figure \ref{fig:geo.strat} visualizes the space $\mathcal B$ as a plane, where each individual point represents an augmentation function. Let $\mathcal C$ denote the linear subspace of $\mathcal B$ consisting of functions of $S$; this is illustrated in Figure \ref{fig:geo.strat} as the horizontal line passing through 0. Let $\mathcal C^{\perp}$ denote the orthogonal complement of $\mathcal C$ in $\mathcal B$, consisting of elements of $\mathcal B$ that are orthogonal to (i.e., have zero inner product with) all elements of $\mathcal C$. As a basic fact in functional analysis, the difference $b-b_{\text{opt}}$ can be decomposed as
$$
b-b_{\text{opt}}=\Pi(b-b_{\text{opt}}|\mathcal C)+\Pi(b-b_{\text{opt}}|\mathcal C^{\perp}),
$$
where $\Pi$ denotes projection (e.g., $\Pi(b-b_{\text{opt}}|\mathcal C)$ is the unique point in $\mathcal C$ that is closest to $b-b_{\text{opt}}$). Orthogonality implies that
$$
\|b-b_{\text{opt}}\|_2^2=\|\Pi(b-b_{\text{opt}}|\mathcal C)\|_2^2+\|\Pi(b-b_{\text{opt}}|\mathcal C^{\perp})\|_2^2.
$$
Let $\Lambda b=b_{\text{opt}}+\Pi(b-b_{\text{opt}}|\mathcal C^{\perp})$, which is closer to $b_{\text{opt}}$ than $b$ is because
$$
\|\Lambda b-b_{\text{opt}}\|_2=\|\Pi(b-b_{\text{opt}}|\mathcal C^{\perp})\|_2\le\|b-b_{\text{opt}}\|_2.
$$
Figure \ref{fig:geo.strat} illustrates that $\Lambda b$ is the result of projecting $b$ into the affine subspace $b_{\text{opt}}+\mathcal C^{\perp}$, which is shown in Figure \ref{fig:geo.strat} as the vertical line passing through $b_{\text{opt}}$. In other words, the operator $\Lambda$ moves a given $b\in\mathcal B$ in parallel to $\mathcal C$ to minimize its distance from $b_{\text{opt}}$. Its relationship with stratified randomization is given in the next result.

\begin{prop}\label{rst:geo.sr}
For any $b\in\mathcal B$, we have
\begin{equation}\label{strat.proj}
\sigma_{\text{st}}^2(b)=\sigma^2(\Lambda b)=\sigma^2(b_{\text{opt}})+\pi(1-\pi)\left\|\Pi\left(b-b_{\text{opt}}|\mathcal C^{\perp}\right)\right\|_2^2.
\end{equation}
\end{prop}

\noindent Thus, for an estimator with (asymptotic) augmentation function $b$, stratified randomization amounts to replacing $b$ with $\Lambda b$ under simple randomization, equating $\sigma_{\text{st}}^2(b)$ to $\sigma^2(\Lambda b)$. Because $\Lambda b$ is closer to $b_{\text{opt}}$ than $b$ is, Proposition \ref{rst:geo.cv} further implies than $\sigma^2(\Lambda b)\le\sigma^2(b)$.

\subsection{Statistical Implications}\label{sec:stat.impl}

\subsubsection{Efficient estimation}\label{sec:eff.est}

Equation \eqref{strat.proj} makes it clear that $\sigma_{\text{st}}^2(b)$ reaches its minimum, $\sigma^2(b_{\text{opt}})$, when and only when $b-b_{\text{opt}}\in\mathcal C$. In other words, $\sigma_{\text{st}}^2(b)$ is minimized by taking $b(\bmW)=b_{\text{opt}}(\bmW)+c(S)$ for an arbitrary function $c$. This result indicates that stratified randomization provides some latitude in efficient estimation by allowing $b$ to differ from $b_{\text{opt}}$ by a function of $S$. More importantly, it confirms that the efficient estimation strategy based on $\widehat\delta_{\text{aug}}(\widehat b)$, with $\widehat b$ targeting $b_{\text{opt}}$, remains appropriate under stratified randomization.

The fact that $\sigma_{\text{st}}^2(b)$ and $\sigma^2(b)$ have the same minimum value is consistent with the main result of \citet{r23}, which states that, for estimating the average treatment effect, the same efficiency bound holds under CAR and simple randomization. Thus, at least for $\delta=\mu_1-\mu_0$, the class of augmented estimators is large enough to attain the efficiency bound under stratified randomization, even though it is not known (by the author) to contain (up to asymptotic equivalence) all regular, asymptotically linear estimators under stratified randomization.

\subsubsection{Analytical analogue of stratified randomization}\label{sec:aa.sr}
Suppose we take $\widehat\delta_{\text{aug}}(b)$ as an initial estimator and seek to improve its efficiency under simple randomization by incorporating information from $S$. Under the augmentation approach, this can be done by subtracting from $\widehat\delta_{\text{aug}}(b)$ an augmentation term based on $S$, leading to
$$
\widehat\delta_{\text{aug}}(b)-\frac1n\sum_{i=1}^n(A_i-\pi)c(S_i)=\widehat\delta_{\text{emp}}-\frac1n\sum_{i=1}^n(A_i-\pi)\{b(\bmW_i)+c(S_i)\}=\widehat\delta_{\text{aug}}(b+c),
$$
with asymptotic variance $\sigma^2(b+c)$ under simple randomization. According to Proposition \ref{rst:geo.cv},
$$
\sigma^2(b+c)=\sigma^2(b_{\text{opt}})+\pi(1-\pi)\|b+c-b_{\text{opt}}\|_2^2,
$$
and its minimum value over $c$ is exactly equal to $\sigma_{\text{st}}^2(b)$:
\begin{equation}\label{strat.var}\begin{aligned}
\min_{c\in\mathcal C}\sigma^2(b+c)&=\sigma^2(b_{\text{opt}})+\pi(1-\pi)\min_{c\in\mathcal C}\|b+c-b_{\text{opt}}\|_2^2\\
&=\sigma^2(b_{\text{opt}})+\pi(1-\pi)\left\|\Pi\left(b-b_{\text{opt}}|\mathcal C^{\perp}\right)\right\|_2^2=\sigma_{\text{st}}^2(b),
\end{aligned}\end{equation}
where the second step follows from geometry and the last one from equation \eqref{strat.proj}. Thus, for a generic augmented estimator, stratified randomization is asymptotically equivalent to attaching an optimal augmentation term based on $S$ under simple randomization.

\subsubsection{Asymptotic equivalence between stratified randomization and analysis}\label{sec:asymp.eq}
Consider the empirical estimator $\widehat\delta_{\text{emp}}=\widehat\delta_{\text{aug}}(0)$. With $b=0$, equation \eqref{strat.var} implies that
\begin{equation}\label{strat.var.emp}
\sigma_{\text{st}}^2(0)=\min_{c\in\mathcal C}\sigma^2(c).
\end{equation}
Minimizing $\sigma^2(c)$ over $c\in\mathcal C$ is a special case of the efficient estimation problem considered in Section \ref{sec:cov.adj}, with $\bmW$ replaced by $S$. Because $S$ is discrete, the minimum of $\sigma^2(c)$ is attained by the regression estimator (described in Appendix A) based on model \eqref{out.reg} with $\bmV=(I(S=2),\dots,I(S=K))^{\text{T}}$. With this choice of $\bmV$, model \eqref{out.reg} is saturated and trivially correct, leading to a regression estimator that is efficient for $S$, which we denote by $\widehat\delta_{\text{reg}}^S$. Now, equation \eqref{strat.var.emp} indicates that the asymptotic variance of $\widehat\delta_{\text{emp}}$ under stratified randomization is equal to the asymptotic variance of $\widehat\delta_{\text{reg}}^S$ under simple randomization. This result coincides with Corollary 2 of \citet{y22} when model \eqref{out.reg} has an identity link. This asymptotic equivalence makes intuitive sense. Stratified randomization produces balanced strata by design, while a stratified analysis attempts to balance covariates by analysis. When the latter is optimally conducted as in $\widehat\delta_{\text{reg}}^S$, it is expected to have the same asymptotic effect as stratified randomization based on $S$.

It should be noted that this asymptotic equivalence between design and analysis is limited to the use of strata information. The stratified analysis based on $\widehat\delta_{\text{reg}}^S$ is a special case of covariate adjustment, which is more general and able to incorporate information from additional baseline covariates in the analysis stage. In contrast, stratified randomization is constrained by the (typically small) number of strata that can be accommodated.

\subsection{Practical Implications}\label{sec:prac.impl}

\subsubsection{Trial design}\label{sec:design}

In designing a trial with stratified randomization, one often faces difficult choices in terms of which covariates to include in the stratification and how many strata to use. Including more covariates increases the potential benefit of stratified randomization, but may result in a large number of strata, some of which may be small and difficult to balance. Some biomarkers may be time-consuming to ascertain, and their inclusion in stratified randomization may cause a delay in randomization and treatment. Fortunately, equation \eqref{strat.var} indicates that the efficiency benefit of stratified randomization can be recovered through covariate adjustment. Therefore, it is desirable but not essential to include all important covariates in the stratification.

To illustrate this point, we give some examples of covariate-adjusted estimators that have the same asymptotic variance under simple versus stratified randomization. The regression estimator $\widehat\delta_{\text{reg}}$ (described in Appendix A) has this property if the vector $\bmV$ in model \eqref{out.reg} includes the dummy variables $(I(S=2),\dots,I(S=K))$, according to Theorem 2 of \citet{y22} and Corollary 1 of \citet{w23}. (There can, and should, be other covariates in $\bmV$ to improve efficiency.) A generic augmented estimator $\widehat\delta_{\text{aug}}(\widehat b)$ can be modified to have this property using the following calibration technique proposed by \citet{b23}. For $k=1,\dots,K$, define
$$
\widehat c_k=\frac{\sum_{i=1}^nI(A_i=1,S_i=k)\widehat\psi(\bmO_i)}{\sum_{i=1}^nI(A_i=1,S_i=k)}
-\frac{\sum_{i=1}^nI(A_i=0,S_i=k)\widehat\psi(\bmO_i)}{\sum_{i=1}^nI(A_i=0,S_i=k)}
-\frac{\sum_{i=1}^nI(S_i=k)\widehat b(\bmW_i)}{\sum_{i=1}^nI(S_i=k)},
$$
where
$$
\widehat\psi(\bmO)=g'(\overline Y_1)\frac{A(Y-\overline Y_1)}{\pi}-
g'(\overline Y_0)\frac{(1-A)(Y-\overline Y_0)}{1-\pi}.
$$
Let $\widehat c(S)=\sum_{k=1}^K\widehat c_kI(S=k)$; then $\widehat\delta_{\text{aug}}(\widehat b+\widehat c)$ is the calibrated version of $\widehat\delta_{\text{aug}}(\widehat b)$ with the desired property. (We note that $\widehat c$ estimates the function $c$ that attains the minima in equation \eqref{strat.var}, where $b$ is the limit of $\widehat b$.)

\subsubsection{Data analysis}\label{sec:analysis}

It is well recognized that stratified randomization should be acknowledged in the statistical analysis of trial data, which is usually interpreted to mean that treatment effect estimation should adjust for the stratification factor as a covariate \citep{ich98,ema03,ema15}. The recent FDA guidance on covariate adjustment \citep{fda23} recommends that, under stratified randomization, \lq\lq A covariate adjustment model should generally include strata variables and can also include covariates not used for stratifying randomization." Accordingly, we will evaluate and compare the potential benefits of adjusting for the stratification factor and possibly other baseline covariates under stratified randomization.

Let us start with the common practice of adjusting only for the stratification factor $S$. There are different ways to adjust for $S$, some of which may be more efficient than others. We will focus on the regression estimator $\widehat\delta_{\text{reg}}^S$, which makes optimal adjustment for $S$, as noted in Section \ref{sec:asymp.eq}. This implies that $\widehat\delta_{\text{reg}}^S$ is asymptotically equivalent to $\widehat\delta_{\text{aug}}(c_{\text{opt}})$, where $c_{\text{opt}}=\argmin_{c\in\mathcal C}\sigma^2(c)$ is the optimal augmentation function based on $S$ only (see Figure \ref{fig:geo.strat} for an illustration). Under simple randomization, $\widehat\delta_{\text{reg}}^S$ is generally more efficient than $\widehat\delta_{\text{emp}}=\widehat\delta_{\text{aug}}(0)$ because $\sigma^2(c_{\text{opt}})=\min_{c\in\mathcal C}\sigma^2(c)\le\sigma^2(0)$. Under stratified randomization, equation \eqref{strat.var} implies that $\widehat\delta_{\text{emp}}$ has asymptotic variance $\sigma_{\text{st}}^2(0)=\min_{c\in\mathcal C}\sigma^2(0+c)=\sigma^2(c_{\text{opt}})$ and that $\widehat\delta_{\text{reg}}^S$ has asymptotic variance $\sigma_{\text{st}}^2(c_{\text{opt}})=\min_{c\in\mathcal C}\sigma^2(c_{\text{opt}}+c)=\sigma^2(c_{\text{opt}})$. Thus, under stratified randomization, $\widehat\delta_{\text{reg}}^S$ is as (in)efficient as $\widehat\delta_{\text{emp}}$ is. This is not surprising because the information in $S$ has already been used (optimally, according to equation \eqref{strat.var}) in the randomization procedure. Only when we incorporate additional information from $\bmW$ into the analysis can we hope to improve efficiency over $\widehat\delta_{\text{emp}}$ under stratified randomization.

The information in $\bmW$ can be incorporated using a regression estimator or an augmented estimator described in Section \ref{sec:cov.adj}. When all of $\bmW$ is included in covariate adjustment, the minimum attainable asymptotic variance is $\sigma^2(b_{\text{opt}})=\min_{b\in\mathcal B}\sigma^2(b)$, as noted in Section \ref{sec:eff.est}. Because $\mathcal B\supset\mathcal C$, this minimum is generally smaller than $\min_{c\in\mathcal C}\sigma^2(c)=\sigma^2(c_{\text{opt}})$, the minimum asymptotic variance that could be attained by adjusting for $S$ alone. The two minima are equal only when $\bmW$ contains no new prognostic information beyond $S$, which is unlikely given how $\bmW$ is chosen in the first place. Of course, $\sigma^2(b_{\text{opt}})$ is only a minimum and not necessarily the actual asymptotic variance of the estimator in use. However, in our numerical experience, it has been difficult to find a situation in which an estimator that adjusts for all of $\bmW$ underperforms an estimator based on $S$, say $\widehat\delta_{\text{reg}}^S$.

As noted in Section \ref{sec:eff.est}, under both simple and stratified randomization, the key to efficient estimation is to adjust for all of $\bmW$ using an appropriate model or method with a limiting augmentation function close to $b_{\text{opt}}$. From this perspective, it is not clear why, under stratified randomization, the stratification factor should take precedence over other baseline covariates in covariate adjustment as suggested by regulatory guidelines \citep{ich98,ema03,ema15,fda23}. This is not to argue against adjusting for $S$ but rather to question the rationale for prioritizing $S$ over other covariates. It is possible that, for a particular model or method, adding dummy variables for $S$ to the covariate vector $\bmW$ may actually improve estimation precision. If this happens, it is because the model or method for covariate adjustment does not make effective use of $\bmW$ (to which $S$ adds no new information), not because $S$ is the stratification factor. If anything, the use of $S$ in stratified randomization should lead to a de-prioritization of $S$ in covariate adjustment, because stratified randomization already makes use of some of the information in $S$, as shown in equation \eqref{strat.var}.

The variance reduction effect of stratified randomization implies that variance estimators developed for simple randomization may be biased upward under stratified randomization, unless the point estimator of $\delta$ has the same asymptotic variance under simple versus stratified randomization. Section \ref{sec:design} gives examples of estimators that do not receive a variance reduction from stratified randomization, including a calibrated estimator that turns out useful for variance estimation. For a generic augmented estimator $\widehat\delta_{\text{aug}}(\widehat b)$ with $\widehat b\to b$, let $\widehat\delta_{\text{aug}}(\widehat b+\widehat c)$ be the calibrated estimator defined in Section \ref{sec:design}; then $\sigma_{\text{st}}^2(b)$ is the common asymptotic variance of both $\widehat\delta_{\text{aug}}(\widehat b)$ and $\widehat\delta_{\text{aug}}(\widehat b+\widehat c)$ under stratified randomization, and is consistently estimated by
$$
\frac1n\sum_{i=1}^n\left[\widehat\psi(\bmO_i)-(A_i-\pi)\{\widehat b(\bmW_i)+\widehat c(S_i)\}\right]^2.
$$
If $\widehat b$ is obtained using a machine learning method, the above variance estimator may suffer from a re-substitution bias in finite samples, which can be removed using cross-validation (see Section 2.3 of \citet{zm19}). For a non-augmented estimator, one may appeal to its asymptotic equivalence to an augmented estimator (say $\widehat\delta_{\text{aug}}(b)$), find a consistent estimator of $b$ (say $\widehat b$), and then apply the calibration technique to obtain a consistent variance estimator under stratified randomization. 

Interestingly, variance estimation provides one possible reason for including $S$ in covariate adjustment. As noted in Section \ref{sec:design}, the regression estimator $\widehat\delta_{\text{reg}}$ does not require a bias correction in variance estimation due to stratified randomization if the covariate vector $\bmV$ in the model includes dummy variables for $S$. It is thus convenient for variance estimation to utilize such a regression estimator with such a covariate configuration. This convenience may be of practical importance but seems insufficient to support the recommendation of always adjusting for $S$ as a profound general principle, because the convenience argument has some limitations. First, it is limited to a specific type of regression estimators and not applicable to general (and potentially more efficient) augmented estimators. Second, as noted in the preceding paragraph, it is straightforward to obtain a valid variance estimator for a generic treatment effect estimator, and computational software is also publicly available \citep{b24}. In light of these limitations, it remains unclear what the rationale is for the general recommendation of always adjusting for $S$ in the analysis following stratified randomization.

\section{Simulation}\label{sim}

To complement the preceding discussion based on asymptotic theory, this section reports a simulation study that evaluates the impact of covariate adjustment and stratified randomization in finite samples. In this simulation study, $\bmW=(W_1,W_2,W_3)^{\text{T}}$ follows the trivariate standard normal distribution with zero correlations, and $S=s(\bmW)=1+I(W_1>0)+2I(W_2>0)$ defines four equal-sized strata based on the signs of $W_1$ and $W_2$. With $\pi=1/2$, treatment assignment is either completely random (as in simple randomization) or stratified by $S$ with a fixed block size of four. The outcome $Y$ may be continuous or binary, and its dependence on $(\bmW,A)$ may be described as follows:
\begin{description}
\item{Scenario A: } $g^{\dagger}\{\epn(Y|\bmW,A)\}=-1+W_1-W_2+W_3+A(1+W_2-2^{-1}W_3)$;
\item{Scenario B: } $g^{\dagger}\{\epn(Y|\bmW,A)\}=-1+W_1-W_2+W_1W_3-W_2W_3+A(1+W_1W_2)$;
\item{Scenario C: } $g^{\dagger}\{\epn(Y|\bmW,A)\}=-1+W_1+W_1^2-(W_2\vee0)^2+A(2+W_3-W_3^2)$;
\item{Scenario D: } $g^{\dagger}\{\epn(Y|\bmW,A)\}=-1+\sqrt{W_1^2+W_2^2}-|W_3|+A$,
\end{description}
where $\vee$ denotes maximum and $g^{\dagger}$ is either the identity function (for continuous $Y$) or the logit function (for binary $Y$). In the continuous case, $Y$ is generated as $Y=\epn(Y|\bmW,A)+\varepsilon$, where $\varepsilon\sim N(0,1)$ independently of $(\bmW,A)$. Scenario A follows a typical generalized linear model structure with interactions between $A$ and $\bmW$. Scenarios B and C feature additional interactions and non-linearity, and Scenario D represents further departure from the typical structure. In each scenario, 2,500 trials are simulated for each type of outcome (continuous or binary) and each randomization scheme (simple or stratified). Each simulated trial consists of $n=200$ or 500 subjects.

Let us first consider the case where $Y$ is continuous. In this case, the estimand $\delta$ is the mean difference between $Y(1)$ and $Y(0)$, and its true value is equal to 1 in each of the four scenarios described earlier. We compare six estimators of $\delta$: the empirical estimator $\widehat\delta_{\text{emp}}$, three regression estimators with different choices of $\bmV$, two augmented estimator (with or without calibration) where $\widehat b$ is obtained using a super learner \citep{zm19}. In the regression estimators, $h$ is the identity function (corresponding to ANCOVA) and $\bmV$ may be a vector of dummy variables for $S$, the original covariate vector $\bmW$, or $\bmW$ concatenated with dummy variables for $S$. The super learner in the augmented estimators is a linear combination (with coefficients optimized in five-fold cross-validation) of the following four algorithms: linear regression, linear regression with pairwise interactions, additive model, and recursive partitioning and regression tree. Under simple randomization, variance estimates are obtained using estimated influence functions, with the use of another five-fold cross-validation for the augmented estimators \citep{zm19}. Under stratified randomization, no adjustment in variance estimation is required for the regression estimators that explicitly adjust for $S$, and calibrated variance estimators (see Section \ref{sec:analysis}) are used for the other estimators. The different methods and designs (i.e., simple vs stratified randomization) are compared in terms of empirical bias, standard deviation, median standard error, and empirical coverage of 95\% confidence intervals. The empirical method combined with simple randomization serves as a reference in efficiency comparisons. The relative efficiency of any method-design combination is defined as the inverse ratio of its empirical variance to that of the reference combination.

Tables \ref{sim.rst.cont.200} and \ref{sim.rst.cont.500} report simulation results for continuous $Y$ with $n=200$ or 500, respectively. The results indicate that estimation bias is negligible (relative to variability), that sampling variability is estimated accurately, and that empirical coverage is adequate, for any method-design combination. Therefore, we will focus on efficiency in comparing different estimators and designs. Under simple randomization, the empirical estimator typically has the lowest efficiency, and the other estimators may or may not bring a substantial improvement, depending on the scenario. The regression estimator adjusting for $S$ only improves efficiency if S is associated with $Y$, as is clearly the case in Scenarios A and B but not in Scenario D. The regression estimators that adjust for $\bmW$ (with or without $S$) improve efficiency (by a larger amount than that achieved by adjusting for $S$ only) in Scenarios A--C but not in Scenario D (where the ANCOVA models are severely mis-specified). The augmented estimators perform similarly to the regression estimators that adjust for $\bmW$ in Scenario A (where the ANCOVA models are correct) and outperform all other estimators in Scenarios B--D. Calibration has no appreciable impact on the super learner based augmentation method. Similar patterns are observed under stratified randomization with one notable exception: adjusting for $S$ only is no longer beneficial, which should be expected because stratified randomization has already made use of stratum information. In terms of utilizing stratum information, stratified randomization seems slightly more effective than covariate adjustment (with the regression estimator based on $S$) in small samples (e.g., $n=200$), although the difference is small and vanishes with increasing $n$. For the regression and augmented estimators based on $\bmW$, it is theoretically possible that stratified randomization may reduce the asymptotic variance, but simulation results indicate that such an improvement tends to be small in magnitude. This is consistent with the observation that calibration brings little improvement to the uncalibrated augmented estimator (under either randomization scheme). Including all of $\bmW$ in covariate adjustment (through regression or, preferably, augmentation) appears to be the main driver of estimation efficiency. When all of $\bmW$ is included in covariate adjustment (through regression or augmentation), it seems difficult to achieve further improvement by making another use of the information in $S$---be it through regression, calibration, or stratified randomization.

In the case of binary $Y$, the estimand is taken to be the log odds ratio, and its true value is found numerically to vary between 0.66 and 0.86 in the four scenarios considered. We compare six estimation methods that are analogous to those compared in the continuous case with the following modifications: the link function has become the logit function, and the ANCOVA models have been replaced by logistic regression models. Except for those changes, the methods are implemented and compared in the same manner as before. Tables \ref{sim.rst.bin.200} and \ref{sim.rst.bin.500} report simulation results for binary $Y$ with $n=200$ or 500, respectively. The results in Tables \ref{sim.rst.bin.200} and \ref{sim.rst.bin.500} follow the same patterns as those in Tables \ref{sim.rst.cont.200} and \ref{sim.rst.cont.500}, although in the binary case efficiency gains tend to be smaller in magnitude.

\section{Example}\label{example}

We now illustrate our findings by analyzing real data from a National Institute of Neurological Disorders and Stroke (NINDS) trial of recombinant tissue plasminogen activator (rt-PA) for treating acute ischemic stroke \citep{n95}. This trial enrolled 624 patients with acute ischemic stroke at eight sites, randomized them in a $1:1$ ratio to rt-PA or placebo within three hours of stroke onset, and evaluated clinical outcomes at 24 hours and three months post-treatment. This trial consisted of two (temporally ordered) parts, with 291 patients enrolled in part 1 and 333 in part 2. The main difference between the two parts was that the time from stroke onset to treatment (TOT) was within 90 minutes in part 1 and between 90 minutes and 180 minutes in part 2. Except for that difference, patients in the two parts were treated and evaluated in the same manner. Within each part, the randomization was stratified by site using permuted blocks of various sizes. In our retrospective analysis, we combine the two parts and consider the overall randomization scheme as stratified by site and study part. Our analysis excludes one site which enrolled only one patient (while the other sites enrolled at least 39 patients per site).

Following previous analyses of the NINDS trial data \citep{n95,i04}, we compare treatments with respect to clinical outcomes at three months, as measured by the Barthel Index, the Modified Rankin Scale, the Glasgow Outcome Scale, and the National Institutes of Health Stroke Scale (NIHSS). The Barthel Index measures the ability to perform activities of daily living and yields a score that ranges from 0 to 100 (complete independence). The Modified Rankin Scale is a simplified overall assessment of function, with a score that ranges from 0 (absence of symptoms) to 5 (severe disability). The Glasgow Outcome Scale is a global assessment of function, with a score that ranges from 1 (good recovery) to 5 (death). The NIHSS is a 42-point scale with 0 indicating no stroke symptoms and higher scores for more severe symptoms. For ease of interpretation and comparison, each original score is linearly re-scaled to the unit interval in such a way that higher scores represent desirable outcomes. For each instrument, the efficacy of rt-PA relative to placebo is measured by the between-group difference (rt-PA minus placebo) in the re-scaled mean score.

In our analysis, the covariate vector $\bmW$ consists of dummy variables for site, continuous TOT, baseline NIHSS, age, preexisting disability, and history of diabetes. (Study part is not included in $\bmW$ because it is determined by TOT.) Six patients with missing covariate data are excluded from this analysis. The remaining data is analyzed using the empirical method, three regression methods based on $S$ and/or $\bmW$, and the augmentation method based on $\bmW$ with or without calibration. The methods are identical to those compared in the simulation study (for a continuous outcome), except that the super learner library includes random forest in addition to the algorithms described in Section \ref{sim}.

Table \ref{ana.rst.ninds} shows point estimates of re-scaled mean score differences together with standard errors that do or do not account for stratified randomization. Accounting for stratified randomization results in minimal changes in the standard errors, suggesting that the stratification factor is not strongly prognostic. Adjusting for the other covariates ($\bmW$) leads to larger reductions in the standard errors. Taken together, the results in Table \ref{ana.rst.ninds} clearly indicate that the use of rt-PA has a beneficial effect, which is statistically significant at the one-sided 2.5\% level (without adjusting for multiplicity) for all scores but NIHSS.

\section{Discussion}\label{disc}

With focus on statistical efficiency, this article highlights a connection between covariate adjustment and stratified randomization with important statistical and practical implications. Specifically, stratified randomization can be regarded as a special form of covariate adjustment that makes optimal use of stratum information. Covariate adjustment is more general and able to incorporate information from all baseline covariates associated with clinical outcomes, although its effectiveness depends on the appropriateness of the method. In designing a trial with stratified randomization, it is not essential to include all important covariates in the stratification, because their prognostic information can be incorporated through covariate adjustment. In the analysis of trial data following stratified randomization, adjusting for the stratification factor only provides no efficiency benefit, and the key to efficient estimation is incorporating information from all important covariates. These findings are supported by both theoretical and empirical results.

While we do not focus on comparing different methods for covariate adjustment, our simulation results suggest that machine learning methods for covariate adjustment perform similarly to or better than standard methods based on parametric models, depending on the appropriateness of the parametric models. In reality, parametric models are almost certainly mis-specified unless they are saturated, which can happen only when all covariates are discrete and their joint distribution has a small number of possible values (relative to the sample size).

While we do not consider other forms of CAR in this article, we note that our findings are equally applicable to the biased coin design. This is because equation \eqref{strat.var.w23}, the basis for our findings concerning stratified randomization, also holds for the biased coin design, according to \citet{w23}.

While we do not consider hypothesis testing in this article, we note that the large-sample performance of a Wald test is largely driven by the asymptotic properties of the point estimator and the variance estimator used in the test. For example, if the point estimator is consistent and asymptotically normal and the variance estimator is biased upward, the Wald test is expected to be conservative (with a deflated type I error rate) in large samples. Assuming that no bias exists in variance estimation, the power of the test generally increases with the efficiency of the point estimator.

While we have adopted a superpopulation perspective in this article, it is worthwhile to relate our findings to recent developments in covariate adjustment and CAR in the literature on randomization-based causal inference \citep{m13,ld20,ly20}, where study subjects (together with their potential outcomes and covariate values) are considered fixed and randomized treatment assignment is the only source of randomness. These authors study unadjusted, stratified and ANCOVA-type estimators of the sample average treatment effect but not general augmented estimators. The CAR mechanisms they consider are not as relevant to clinical trials as the stratified randomization mechanism considered in this article. Specifically, the rerandomization design considered by \citet{ld20} is infeasible for clinical trials as it assumes that covariate data are available for all patients before randomization occurs, and the stratified randomization design in \citet{m13} and \citet{ly20} assumes that each stratum is randomized as one block, which is rarely the case in clinical trials. Nonetheless, there are interesting similarities between their findings and ours. For example, Theorem 4 of \citet{m13}, a finite-sample result indicating that the variance of the stratified estimator is similar but slightly lower under stratified versus simple randomization, is consistent with our asymptotic theory and helps explain the empirical observation that stratified randomization seems slightly more effective than stratified analysis at $n=200$. The observation in Section 4.1.1 of \citet{ld20}, that rerandomization does not change the optimally adjusted estimator or its asymptotic distribution, is consistent with our finding in Section \ref{sec:eff.est} that $\widehat\delta_{\text{aug}}(b_{\text{opt}})$ is optimal under both simple and stratified randomization with the same asymptotic variance. The asymptotic near equivalence between rerandomization and optimal adjustment, noted in Section 4.1.2 of \citet{ld20}, is similar in spirit to our asymptotic equivalence result in Section \ref{sec:asymp.eq}.

\section*{Appendix A: Regression Estimators}

A regression (or G-computation) estimator of $\delta$ may be obtained by fitting a standard regression model:
\begin{equation}\label{out.reg}\tag{A.1}
\epn(Y|A,\bmV)=h\left((1,A,\bmV^{\text{T}},A\bmV^{\text{T}})\bmbeta\right),
\end{equation}
where $\bmV$ is a vector-valued function of $\bmW$, $h$ a specified inverse link function, and $\bmbeta$ an unknown parameter vector. The interaction term in model \eqref{out.reg} is optional in practice but necessary in the present discussion. Suppose $\bmbeta$ is estimated by solving the estimating equation
\begin{equation}\label{est.eqn}\tag{A.2}
\sum_{i=1}^n\left\{Y_i-h(1,A_i,\bmV_i^{\text{T}},A_i\bmV_i^{\text{T}})\bmbeta\right\}
(1,A_i,\bmV_i^{\text{T}},A_i\bmV_i^{\text{T}})^{\text{T}}=\bmzero.
\end{equation}
This is the likelihood equation for $\bmbeta$ if model \eqref{out.reg} is a generalized linear model with a canonical link function. In general, equation \eqref{est.eqn} is an unbiased estimating equation for $\bmbeta$ for an arbitrary link function and without making distributional assumptions. Let $\widehat\bmbeta$ denote the solution to equation \eqref{est.eqn}; then $\delta$ can be estimated by the regression estimator $\widehat\delta_{\text{reg}}=g(\widehat\mu_{\text{reg},1})-g(\widehat\mu_{\text{reg},0})$, where
$$
\widehat\mu_{\text{reg},a}=\frac1n\sum_{i=1}^nh\left((1,a,\bmV_i^{\text{T}},a\bmV_i^{\text{T}})\widehat\bmbeta\right),\qquad a=0,1.
$$
Examples of the regression estimator include the analysis of covariance (ANCOVA) estimator for continuous outcomes \citep{t08} and the standardized logistic regression estimator for binary outcomes \citep{m09}. It is worth noting that, whether model \eqref{out.reg} is correct or not, $\widehat\delta_{\text{reg}}$ is consistent for $\delta$ and asymptotically normal with asymptotic variance $\var\{\psi(\bmO)-(A-\pi)b_{\text{reg}}(\bmW)\}$, where
$$
b_{\text{reg}}(\bmW)=\frac{g'(\mu_1)\left\{h\left((1,1,\bmV^{\text{T}},\bmV^{\text{T}})\bmbeta^*\right)-\mu_1\right\}}{\pi}+\frac{g'(\mu_0)\left\{h\left((1,0,\bmV^{\text{T}},0\bmV^{\text{T}})\bmbeta^*\right)-\mu_0\right\}}{1-\pi}
$$
and $\bmbeta^*$ is the limit of $\widehat\bmbeta$. This robustness property has been noted by \citet{m09} and \citet{z19} among others.

\section*{Appendix B: Proofs}

\subsection*{Proof of Proposition \ref{rst:geo.cv}}

It can be shown as in Web Appendix C of \citet{z08} that $(A-\pi)b_{\text{opt}}(\bmW)$ is the projection of $\psi(\bmO)\in L_2(\bmO)$ into the linear subspace $\{(A-\pi)d(\bmW):d\in L_2(\bmW)\}$. Therefore, the residual $\psi(\bmO)-(A-\pi)b_{\text{opt}}(\bmW)\}$ is orthogonal to $(A-\pi)d(\bmW)$ for any $d\in L_2(\bmW)$. In particular, we have
$$
\cov\{\psi(\bmO)-(A-\pi)b_{\text{opt}}(\bmW)\},(A-\pi)(b-b_{\text{opt}})(\bmW)\}=0.
$$
It follows that
\begin{equation*}\begin{aligned}
\sigma^2(b)&=\var\{\psi(\bmO)-(A-\pi)b(\bmW)\}\\
&=\var\{\psi(\bmO)-(A-\pi)b_{\text{opt}}(\bmW)-(A-\pi)(b-b_{\text{opt}})(\bmW)\}\\
&=\var\{\psi(\bmO)-(A-\pi)b_{\text{opt}}(\bmW)\}+\var\{(A-\pi)(b-b_{\text{opt}})(\bmW)\}\\
&=\sigma^2(b_{\text{opt}})+\epn[\{(A-\pi)(b-b_{\text{opt}})(\bmW)\}^2]\\
&=\sigma^2(b_{\text{opt}})+\epn\{(A-\pi)^2\}\epn\{(b-b_{\text{opt}})(\bmW)^2\}\\
&=\sigma^2(b_{\text{opt}})+\pi(1-\pi)\|b-b_{\text{opt}}\|_2^2.
\end{aligned}\end{equation*}

\subsection*{Proof of Proposition \ref{rst:geo.sr}}

Let us first reproduce equation \eqref{strat.var.w23}:
$$
\sigma_{\text{st}}^2(b)=\sigma^2(b)
-\frac{1}{\pi(1-\pi)}\epn\left\{\left(\epn[(A-\pi)\{\psi(\bmO)-(A-\pi)b(\bmW)\}|S]\right)^2\right\}.
$$
We note that
\begin{equation*}\begin{aligned}
&\quad\epn[(A-\pi)\{\psi(\bmO)-(A-\pi)b(\bmW)\}|\bmW]\\
&=\epn\{(A-\pi)\psi(\bmO)|\bmW\}-\epn\{(A-\pi)^2b(\bmW)|\bmW\}\\
&=\sum_{a=0}^1\pr(A=a|\bmW)(a-\pi)\epn\{\psi(\bmO)|A=a,\bmW\}-\epn\{(A-\pi)^2\}b(\bmW)\\
&=\pi(1-\pi)\epn\{\psi(\bmO)|A=1,\bmW\}-\pi(1-\pi)\epn\{\psi(\bmO)|A=0,\bmW\}-\pi(1-\pi)b(\bmW)\\
&=\pi(1-\pi)\{b_{\text{opt}}(\bmW)-b(\bmW)\},
\end{aligned}\end{equation*}
where the last step follows from the definition of $b_{\text{opt}}$. Therefore,
\begin{multline*}
\epn[(A-\pi)\{\psi(\bmO)-(A-\pi)b(\bmW)\}|S]\\
=\pi(1-\pi)\epn\{b_{\text{opt}}(\bmW)-b(\bmW)|S\}
=\pi(1-\pi)\Pi(b_{\text{opt}}-b|\mathcal C)(\bmW)
\end{multline*}
and
\begin{equation}\label{term1}\tag{B.1}
\sigma_{\text{st}}^2(b)-\sigma^2(b)=-\pi(1-\pi)\|\Pi(b_{\text{opt}}-b|\mathcal C)\|_2^2
=-\pi(1-\pi)\|\Pi(b-b_{\text{opt}}|\mathcal C)\|_2^2.
\end{equation}
It follows immediately from equation \eqref{geo.cv} that
\begin{equation}\label{term2}\tag{B.2}
\sigma^2(b)-\sigma^2(b_{\text{opt}})=\pi(1-\pi)\|b-b_{\text{opt}}\|_2^2.
\end{equation}
Adding \eqref{term1} and \eqref{term2} yields
\begin{equation*}
\sigma_{\text{st}}^2(b)-\sigma^2(b_{\text{opt}})=\pi(1-\pi)\left(\|b-b_{\text{opt}}\|_2^2-\|\Pi(b-b_{\text{opt}}|\mathcal C)\|_2^2\right)
=\pi(1-\pi)\left\|\Pi(b-b_{\text{opt}}|\mathcal C^{\perp})\right\|_2^2.
\end{equation*}

\renewcommand{\baselinestretch}{1.1}
\begin{table}[htbp]
\caption{Simulation results for continuous $Y$ and $n=200$: empirical bias, standard deviation (SD), relative efficiency (RE), median standard error (SE) and coverage probability (CP) for estimating the mean difference using the emp(irical) method, the reg(ression) method based on $S$, $\bmW$ or both, and the aug(mentation) method without or with cal(ibration).}\label{sim.rst.cont.200}
\newcolumntype{d}{D{.}{.}{2}}
\newcolumntype{e}{D{.}{.}{1}}
\begin{center}
\begin{tabular}{ccddeddcddedd}
\hline
\hline
Scenario&Method&\multicolumn{5}{c}{Simple Randomization}&&\multicolumn{5}{c}{Stratified Randomization}\\
\cline{3-7}\cline{9-13}
&&\multicolumn{1}{c}{Bias}&\multicolumn{1}{c}{SD}&\multicolumn{1}{c}{RE}&\multicolumn{1}{c}{SE}&\multicolumn{1}{c}{CP}&&\multicolumn{1}{c}{Bias}&\multicolumn{1}{c}{SD}&\multicolumn{1}{c}{RE}&\multicolumn{1}{c}{SE}&\multicolumn{1}{c}{CP}\\
\hline
A&emp&-0.01&0.25&1.0&0.25&0.94&&-0.01&0.21&1.5&0.21&0.95\\
&reg ($S$)&-0.01&0.22&1.3&0.22&0.96&&-0.01&0.21&1.5&0.22&0.96\\
&reg ($\bmW$)&-0.01&0.16&2.4&0.17&0.96&&-0.01&0.16&2.5&0.16&0.96\\
&reg ($\bmW+S$)&-0.01&0.16&2.4&0.17&0.96&&-0.01&0.16&2.5&0.17&0.96\\
&aug&-0.01&0.16&2.4&0.17&0.95&&-0.01&0.16&2.5&0.17&0.96\\
&aug\_cal&-0.01&0.16&2.4&0.17&0.95&&-0.01&0.16&2.5&0.17&0.96\\
\hline
B&emp&0.01&0.33&1.0&0.33&0.94&&-0.01&0.29&1.3&0.28&0.95\\
&reg ($S$)&0.00&0.29&1.3&0.29&0.96&&-0.01&0.29&1.3&0.29&0.95\\
&reg ($\bmW$)&0.00&0.27&1.5&0.27&0.95&&-0.01&0.26&1.6&0.26&0.95\\
&reg ($\bmW+S$)&0.00&0.27&1.5&0.27&0.96&&-0.01&0.26&1.6&0.27&0.96\\
&aug&0.00&0.16&4.2&0.16&0.95&&0.00&0.16&4.3&0.16&0.95\\
&aug\_cal&0.00&0.16&4.2&0.16&0.95&&0.00&0.16&4.3&0.16&0.95\\
\hline
C&emp&0.01&0.37&1.0&0.36&0.95&&0.00&0.35&1.1&0.34&0.94\\
&reg ($S$)&0.00&0.34&1.1&0.35&0.95&&0.00&0.35&1.1&0.35&0.95\\
&reg ($\bmW$)&0.01&0.31&1.4&0.32&0.95&&0.01&0.32&1.3&0.31&0.94\\
&reg ($\bmW+S$)&0.01&0.31&1.4&0.32&0.96&&0.01&0.32&1.3&0.32&0.95\\
&aug&0.03&0.22&2.7&0.22&0.95&&0.02&0.22&2.8&0.22&0.95\\
&aug\_cal&0.03&0.22&2.7&0.22&0.94&&0.02&0.22&2.8&0.22&0.95\\
\hline
D&emp&0.00&0.19&1.0&0.19&0.95&&0.00&0.19&0.9&0.19&0.94\\
&reg ($S$)&0.00&0.19&1.0&0.20&0.96&&0.00&0.19&0.9&0.19&0.95\\
&reg ($\bmW$)&0.00&0.19&1.0&0.20&0.96&&0.00&0.19&0.9&0.19&0.95\\
&reg ($\bmW+S$)&0.00&0.19&1.0&0.20&0.96&&0.00&0.19&0.9&0.20&0.95\\
&aug&0.00&0.15&1.5&0.16&0.96&&0.00&0.15&1.5&0.15&0.95\\
&aug\_cal&0.00&0.15&1.5&0.15&0.96&&0.00&0.15&1.5&0.15&0.95\\
\hline
\end{tabular}
\end{center}
\end{table}

\renewcommand{\baselinestretch}{1.1}
\begin{table}[htbp]
\caption{Simulation results for continuous $Y$ and $n=500$: empirical bias, standard deviation (SD), relative efficiency (RE), median standard error (SE) and coverage probability (CP) for estimating the mean difference using the emp(irical) method, the reg(ression) method based on $S$, $\bmW$ or both, and the aug(mentation) method without or with cal(ibration).}\label{sim.rst.cont.500}
\newcolumntype{d}{D{.}{.}{2}}
\newcolumntype{e}{D{.}{.}{1}}
\begin{center}
\begin{tabular}{ccddeddcddedd}
\hline
\hline
Scenario&Method&\multicolumn{5}{c}{Simple Randomization}&&\multicolumn{5}{c}{Stratified Randomization}\\
\cline{3-7}\cline{9-13}
&&\multicolumn{1}{c}{Bias}&\multicolumn{1}{c}{SD}&\multicolumn{1}{c}{RE}&\multicolumn{1}{c}{SE}&\multicolumn{1}{c}{CP}&&\multicolumn{1}{c}{Bias}&\multicolumn{1}{c}{SD}&\multicolumn{1}{c}{RE}&\multicolumn{1}{c}{SE}&\multicolumn{1}{c}{CP}\\
\hline
A&emp&0.00&0.16&1.0&0.16&0.95&&0.00&0.14&1.3&0.14&0.94\\
&reg ($S$)&0.00&0.14&1.3&0.14&0.96&&0.00&0.14&1.3&0.14&0.95\\
&reg ($\bmW$)&0.00&0.10&2.4&0.10&0.96&&0.00&0.10&2.3&0.10&0.95\\
&reg ($\bmW+S$)&0.00&0.10&2.4&0.10&0.96&&0.00&0.10&2.3&0.10&0.95\\
&aug&0.00&0.10&2.4&0.10&0.95&&0.00&0.10&2.3&0.10&0.95\\
&aug\_cal&0.00&0.10&2.4&0.10&0.96&&0.00&0.10&2.3&0.10&0.95\\
\hline
B&emp&-0.01&0.21&1.0&0.21&0.95&&0.00&0.19&1.3&0.18&0.95\\
&reg ($S$)&-0.01&0.18&1.3&0.18&0.95&&0.00&0.19&1.3&0.18&0.95\\
&reg ($\bmW$)&-0.01&0.17&1.5&0.17&0.95&&0.00&0.17&1.5&0.17&0.95\\
&reg ($\bmW+S$)&-0.01&0.17&1.6&0.17&0.96&&0.00&0.17&1.5&0.17&0.95\\
&aug&0.00&0.10&4.5&0.10&0.96&&0.00&0.10&4.4&0.10&0.95\\
&aug\_cal&0.00&0.10&4.5&0.10&0.96&&0.00&0.10&4.4&0.10&0.95\\
\hline
C&emp&-0.01&0.24&1.0&0.23&0.94&&0.00&0.21&1.2&0.22&0.96\\
&reg ($S$)&-0.01&0.22&1.1&0.22&0.95&&0.00&0.21&1.2&0.22&0.96\\
&reg ($\bmW$)&0.00&0.20&1.4&0.20&0.95&&0.00&0.19&1.5&0.20&0.95\\
&reg ($\bmW+S$)&0.00&0.20&1.4&0.20&0.95&&0.00&0.20&1.5&0.20&0.95\\
&aug&0.01&0.14&2.8&0.14&0.94&&0.01&0.14&3.0&0.14&0.95\\
&aug\_cal&0.01&0.14&2.8&0.14&0.94&&0.01&0.14&3.0&0.14&0.95\\
\hline
D&emp&0.00&0.12&1.0&0.12&0.95&&0.00&0.12&1.0&0.12&0.95\\
&reg ($S$)&0.00&0.12&1.0&0.12&0.95&&0.00&0.12&1.0&0.12&0.95\\
&reg ($\bmW$)&0.00&0.12&1.0&0.12&0.96&&0.00&0.12&1.0&0.12&0.95\\
&reg ($\bmW+S$)&0.00&0.12&1.0&0.12&0.96&&0.00&0.12&1.0&0.12&0.96\\
&aug&0.00&0.10&1.5&0.10&0.95&&0.00&0.10&1.6&0.10&0.95\\
&aug\_cal&0.00&0.10&1.5&0.10&0.95&&0.00&0.10&1.6&0.10&0.95\\
\hline
\end{tabular}
\end{center}
\end{table}

\renewcommand{\baselinestretch}{1.1}
\begin{table}[htbp]
\caption{Simulation results for binary $Y$ and $n=200$: empirical bias, standard deviation (SD), relative efficiency (RE), median standard error (SE) and coverage probability (CP) for estimating the log odds ratio using the emp(irical) method, the reg(ression) method based on $S$, $\bmW$ or both, and the aug(mentation) method without or with cal(ibration).}\label{sim.rst.bin.200}
\newcolumntype{d}{D{.}{.}{2}}
\newcolumntype{e}{D{.}{.}{1}}
\begin{center}
\begin{tabular}{ccddeddcddedd}
\hline
\hline
Scenario&Method&\multicolumn{5}{c}{Simple Randomization}&&\multicolumn{5}{c}{Stratified Randomization}\\
\cline{3-7}\cline{9-13}
&&\multicolumn{1}{c}{Bias}&\multicolumn{1}{c}{SD}&\multicolumn{1}{c}{RE}&\multicolumn{1}{c}{SE}&\multicolumn{1}{c}{CP}&&\multicolumn{1}{c}{Bias}&\multicolumn{1}{c}{SD}&\multicolumn{1}{c}{RE}&\multicolumn{1}{c}{SE}&\multicolumn{1}{c}{CP}\\
\hline
A&emp&0.00&0.29&1.0&0.29&0.95&&0.00&0.28&1.1&0.27&0.95\\
&reg ($S$)&0.00&0.28&1.1&0.28&0.95&&0.00&0.28&1.1&0.27&0.95\\
&reg ($\bmW$)&0.00&0.26&1.3&0.26&0.95&&0.00&0.26&1.3&0.25&0.95\\
&reg ($\bmW+S$)&0.00&0.26&1.2&0.26&0.94&&0.00&0.26&1.3&0.25&0.95\\
&aug&0.00&0.26&1.2&0.27&0.96&&0.00&0.26&1.3&0.27&0.96\\
&aug\_cal&0.00&0.27&1.2&0.27&0.95&&0.00&0.26&1.3&0.27&0.96\\
\hline
B&emp&0.00&0.30&1.0&0.29&0.95&&0.01&0.28&1.1&0.28&0.95\\
&reg ($S$)&0.00&0.29&1.1&0.28&0.95&&0.01&0.28&1.1&0.28&0.95\\
&reg ($\bmW$)&0.00&0.28&1.2&0.27&0.94&&0.00&0.27&1.2&0.27&0.95\\
&reg ($\bmW+S$)&0.00&0.28&1.1&0.27&0.94&&0.00&0.27&1.2&0.27&0.95\\
&aug&0.00&0.27&1.3&0.27&0.95&&0.00&0.26&1.3&0.26&0.96\\
&aug\_cal&0.00&0.27&1.3&0.26&0.95&&0.00&0.26&1.3&0.26&0.96\\
\hline
C&emp&0.00&0.29&1.0&0.29&0.96&&0.00&0.28&1.1&0.28&0.95\\
&reg ($S$)&0.01&0.28&1.1&0.28&0.95&&0.00&0.28&1.1&0.28&0.95\\
&reg ($\bmW$)&0.01&0.27&1.1&0.27&0.95&&0.01&0.27&1.2&0.27&0.95\\
&reg ($\bmW+S$)&0.01&0.28&1.1&0.27&0.94&&0.01&0.27&1.1&0.27&0.95\\
&aug&0.01&0.26&1.3&0.26&0.95&&0.01&0.25&1.3&0.26&0.96\\
&aug\_cal&0.01&0.26&1.3&0.26&0.95&&0.01&0.25&1.3&0.26&0.96\\
\hline
D&emp&0.00&0.29&1.0&0.29&0.95&&0.01&0.29&1.0&0.29&0.95\\
&reg ($S$)&0.00&0.29&1.0&0.29&0.95&&0.01&0.29&1.0&0.29&0.95\\
&reg ($\bmW$)&0.00&0.29&1.0&0.29&0.95&&0.01&0.29&1.0&0.29&0.95\\
&reg ($\bmW+S$)&0.00&0.30&1.0&0.29&0.95&&0.01&0.29&1.0&0.29&0.95\\
&aug&0.00&0.28&1.1&0.28&0.95&&0.00&0.28&1.1&0.28&0.95\\
&aug\_cal&0.00&0.28&1.1&0.28&0.95&&0.00&0.28&1.1&0.28&0.95\\
\hline
\end{tabular}
\end{center}
\end{table}

\renewcommand{\baselinestretch}{1.1}
\begin{table}[htbp]
\caption{Simulation results for binary $Y$ and $n=500$: empirical bias, standard deviation (SD), relative efficiency (RE), median standard error (SE) and coverage probability (CP) for estimating the log odds ratio using the emp(irical) method, the reg(ression) method based on $S$, $\bmW$ or both, and the aug(mentation) method without or with cal(ibration).}\label{sim.rst.bin.500}
\newcolumntype{d}{D{.}{.}{2}}
\newcolumntype{e}{D{.}{.}{1}}
\begin{center}
\begin{tabular}{ccddeddcddedd}
\hline
\hline
Scenario&Method&\multicolumn{5}{c}{Simple Randomization}&&\multicolumn{5}{c}{Stratified Randomization}\\
\cline{3-7}\cline{9-13}
&&\multicolumn{1}{c}{Bias}&\multicolumn{1}{c}{SD}&\multicolumn{1}{c}{RE}&\multicolumn{1}{c}{SE}&\multicolumn{1}{c}{CP}&&\multicolumn{1}{c}{Bias}&\multicolumn{1}{c}{SD}&\multicolumn{1}{c}{RE}&\multicolumn{1}{c}{SE}&\multicolumn{1}{c}{CP}\\
\hline
A&emp&0.00&0.19&1.0&0.18&0.95&&0.00&0.18&1.2&0.17&0.94\\
&reg ($S$)&0.00&0.18&1.1&0.17&0.95&&0.00&0.18&1.2&0.17&0.95\\
&reg ($\bmW$)&0.00&0.17&1.3&0.16&0.95&&0.00&0.16&1.3&0.16&0.95\\
&reg ($\bmW+S$)&0.00&0.17&1.3&0.16&0.95&&0.00&0.16&1.3&0.16&0.95\\
&aug&0.00&0.17&1.3&0.16&0.95&&0.00&0.16&1.3&0.16&0.95\\
&aug\_cal&0.00&0.17&1.3&0.16&0.95&&0.00&0.16&1.3&0.16&0.95\\
\hline
B&emp&0.00&0.18&1.0&0.18&0.95&&0.00&0.18&1.1&0.18&0.95\\
&reg ($S$)&0.00&0.18&1.1&0.18&0.95&&0.00&0.18&1.1&0.18&0.95\\
&reg ($\bmW$)&0.00&0.17&1.1&0.17&0.95&&0.00&0.17&1.1&0.17&0.95\\
&reg ($\bmW+S$)&0.00&0.17&1.1&0.17&0.95&&0.00&0.17&1.1&0.17&0.95\\
&aug&-0.01&0.16&1.3&0.16&0.95&&-0.01&0.16&1.3&0.16&0.95\\
&aug\_cal&-0.01&0.16&1.3&0.16&0.95&&-0.01&0.16&1.3&0.16&0.95\\
\hline
C&emp&0.00&0.19&1.0&0.18&0.94&&0.00&0.18&1.1&0.18&0.95\\
&reg ($S$)&0.00&0.18&1.1&0.18&0.94&&0.00&0.18&1.1&0.18&0.95\\
&reg ($\bmW$)&0.00&0.18&1.1&0.17&0.94&&0.00&0.17&1.2&0.17&0.95\\
&reg ($\bmW+S$)&0.00&0.18&1.1&0.17&0.94&&0.00&0.17&1.2&0.17&0.95\\
&aug&0.01&0.16&1.3&0.16&0.94&&0.01&0.16&1.4&0.16&0.95\\
&aug\_cal&0.01&0.16&1.3&0.16&0.94&&0.01&0.16&1.4&0.16&0.95\\
\hline
D&emp&0.00&0.18&1.0&0.18&0.95&&0.00&0.19&1.0&0.18&0.95\\
&reg ($S$)&0.00&0.18&1.0&0.18&0.95&&0.00&0.19&1.0&0.18&0.95\\
&reg ($\bmW$)&0.00&0.19&1.0&0.18&0.95&&0.00&0.19&1.0&0.18&0.95\\
&reg ($\bmW+S$)&0.00&0.19&1.0&0.18&0.95&&0.00&0.19&1.0&0.18&0.95\\
&aug&0.00&0.17&1.1&0.18&0.95&&0.00&0.18&1.1&0.18&0.95\\
&aug\_cal&0.00&0.17&1.1&0.18&0.95&&0.00&0.18&1.1&0.18&0.95\\
\hline
\end{tabular}
\end{center}
\end{table}

\renewcommand{\baselinestretch}{1.1}
\begin{table}[htbp]
\caption{Results of analyzing the NINDS trial data: point estimates and standard errors (uncorrected or corrected for stratified randomization) for re-scaled mean score differences between rt-PA and placebo, obtained using the emp(irical) method, the reg(ression) method based on $S$, $\bmW$ or both, and the aug(mentation) method based on $\bmW$ with or without cal(ibration).}\label{ana.rst.ninds}
\newcolumntype{d}{D{.}{.}{3}}
\begin{center}
\begin{tabular}{ccddd}
\hline
\hline
Outcome&Method&\multicolumn{1}{c}{Mean}&\multicolumn{2}{c}{Standard Error}\\
\cline{4-5}
&&\multicolumn{1}{c}{Difference}&\multicolumn{1}{c}{Uncorrected}&\multicolumn{1}{c}{Corrected}\\
\hline
Barthel Index&emp&0.083&0.033&0.033\\
&reg ($S$)&0.080&0.034&0.035\\
&reg ($\bmW$)&0.074&0.028&0.028\\
&reg ($\bmW+S$)&0.074&0.029&0.029\\
&aug&0.077&0.028&0.028\\
&aug\_cal&0.076&0.028&0.028\\
\hline
Modified Rankin Scale&emp&0.107&0.033&0.033\\
&reg ($S$)&0.105&0.034&0.034\\
&reg ($\bmW$)&0.094&0.028&0.028\\
&reg ($\bmW+S$)&0.094&0.029&0.029\\
&aug&0.094&0.028&0.027\\
&aug\_cal&0.093&0.027&0.027\\
\hline
Glasgow Outcome Scale&emp&0.075&0.030&0.029\\
&reg ($S$)&0.074&0.030&0.030\\
&reg ($\bmW$)&0.064&0.026&0.025\\
&reg ($\bmW+S$)&0.064&0.026&0.026\\
&aug&0.064&0.025&0.025\\
&aug\_cal&0.064&0.025&0.025\\
\hline
NIHSS&emp&0.052&0.029&0.029\\
&reg ($S$)&0.050&0.030&0.030\\
&reg ($\bmW$)&0.040&0.026&0.026\\
&reg ($\bmW+S$)&0.039&0.027&0.027\\
&aug&0.044&0.025&0.025\\
&aug\_cal&0.043&0.025&0.025\\
\hline
\end{tabular}
\end{center}
\end{table}

\end{document}